\documentclass{icrc29}

\usepackage{graphicx,amssymb,amsmath,times}
\begin{document}

\title[Analysis of extensive air showers with the hybrid code SENECA] {Analysis of extensive air showers with the hybrid
code SENECA}

\author[Jeferson A. Ortiz, Vitor de Souza and Gustavo Medina-Tanco] {Jeferson A. Ortiz, Vitor de Souza and Gustavo Medina-Tanco \\
Instituto de Astronomia, Geof\'{\i}sica e Ci\^encias Atmosf\'ericas, Universidade de S\~ao Paulo, Brasil\\ }

\presenter{Presenter: Jeferson A. Ortiz (jortiz@astro.iag.usp.br), bra-ortiz-JA-abs1-he14-poster}

\maketitle

\begin{abstract}

The ultrahigh energy tail of the cosmic ray spectrum has been
explored with unprecedented detail. For this reason, new
experiments are exerting a severe pressure on extensive air shower
modelling. Detailed fast codes are in need in order to extract and
understand the richness of information now available. In this
sense we explore the potential of SENECA, an efficient hybrid
tridimensional simulation code, as a valid practical alternative
to full Monte Carlo simulations of extensive air showers generated
by ultrahigh energy cosmic rays. We discuss the influence of this
approach on the main longitudinal characteristics of proton and
gamma induced air showers for different hadronic interaction
models. We also show the comparisons of our predictions with those
of CORSIKA code.
\end{abstract}

%
%
\section{Introduction}

Since the very first observations, ultra-high energy cosmic rays (UHECR)
have been an open question and a priority in astroparticle physics.
Their origin, nature and possible acceleration mechanisms are still a
mystery.

Due to the very low flux of high energy cosmic rays, measuring extensive air
showers (EAS) is the only possible technique to learn about the shape of the UHECR
spectrum and their chemical composition. Two different ways have been historically
applied to observe and analyze EAS's: ground array of detectors and optical detectors.
Surface detectors measure a lateral density sample and trigger in coincidence
when charged particles pass through them while optical detectors (i.e., fluorescence
detectors) observe the longitudinal profile evolution by measuring the fluorescence
light from atmospheric nitrogen excitation produced by the ionization of the secondary
charged particles (essentially electrons and positrons).

The combination of shower observables (such as lateral density, the depth of maximum
shower development ($X_{\rm max}$), the number of charged particles at shower maximum
($S_{\rm max}$) and number of muons ($N_\mu$) at detector observation level) and
simulation techniques is the current way to obtain information about the primary energy,
composition and arrival direction. For this purpose the shower simulation should provide
all possible, and ideally the necessary, information to interpret measurements of shower
parameters.

Many shower simulation packages have been developed over the
years. Most of them are based on the Monte Carlo method and
simulate complete high energy showers with well described
fluctuations in the first particle interactions and realistic
distributions of energy of shower particles. Recently, different
ways of calculating the air shower development have been
proposed~\cite{Bossard01,Engel04c,Alvarez02}. Most of them combine
the traditional Monte Carlo scheme with a system of
electromagnetic and hadronic cascade equations or pre-simulated
showers, described with parameterizations.

In the present contribution we analyze the results obtained by the
SENECA~\cite{Drescher03} code which is based on the Monte Carlo
calculation of the first and final stages of the air shower
development, and on a cascade equation system that connects both
stages reproducing the longitudinal shower development. We explore
the main longitudinal shower characteristics of proton and gamma
initiated air showers at ultra-high energy, as predicted by the
QGSJET~\cite{qgsjetb} and SIBYLL~\cite{sibyll} hadronic
interaction models. The SENECA predictions are compared with the
well tested CORSIKA (COsmic Ray SImulations for KAscade)
simulation code~\cite{Heck98a}.

%
%
\section{Air Shower Modelling}

The main goal of this approach is the generation of EAS's in a fast manner, obtaining
the correct description of the fluctuations in showers and giving the average values
for the shower characteristics. Even though the SENECA code describes both longitudinal
and lateral air shower developments, the simulation scheme is used here to generate
large statistics of longitudinal shower profiles applicable mainly to the present
fluorescence detectors, such as Pierre Auger Observatory~\cite{Auger} and
HiRes~\cite{HiRes01a}, as well to the future telescope EUSO~\cite{Catalano01}.

For the present work we track explicitly every particle with energy above the fraction
$f$=$E_{0}/1.000$, where $E_{0}$ is the primary shower energy, studying in detail the
initial part of the shower. All secondary particles with energy below the mentioned
fraction are taken as initial conditions to initialize a system of hadronic and
electromagnetic cascade equations~\cite{Drescher03}. We use the cascade
equations with the minimum electromagnetic energy thresholds of 1~GeV for the
electromagnetic component ($E_{\mathrm {min}}^{\mathrm{em}}$) and
$E_{\mathrm {min}}^{\mathrm{had}}$=$10^{4}$~GeV for the hadronic component.
The hadronic interactions at high energies are calculated with both QGSJET01~\cite{qgsjetb}
and SIBYLL2.1~\cite{sibyll} models. The adopted kinetic energy cutoffs for all simulations were
50~MeV (0.3~MeV) for hadrons and muons (electrons and positrons).
\begin{figure}[ht]
\includegraphics[height=5.0cm]{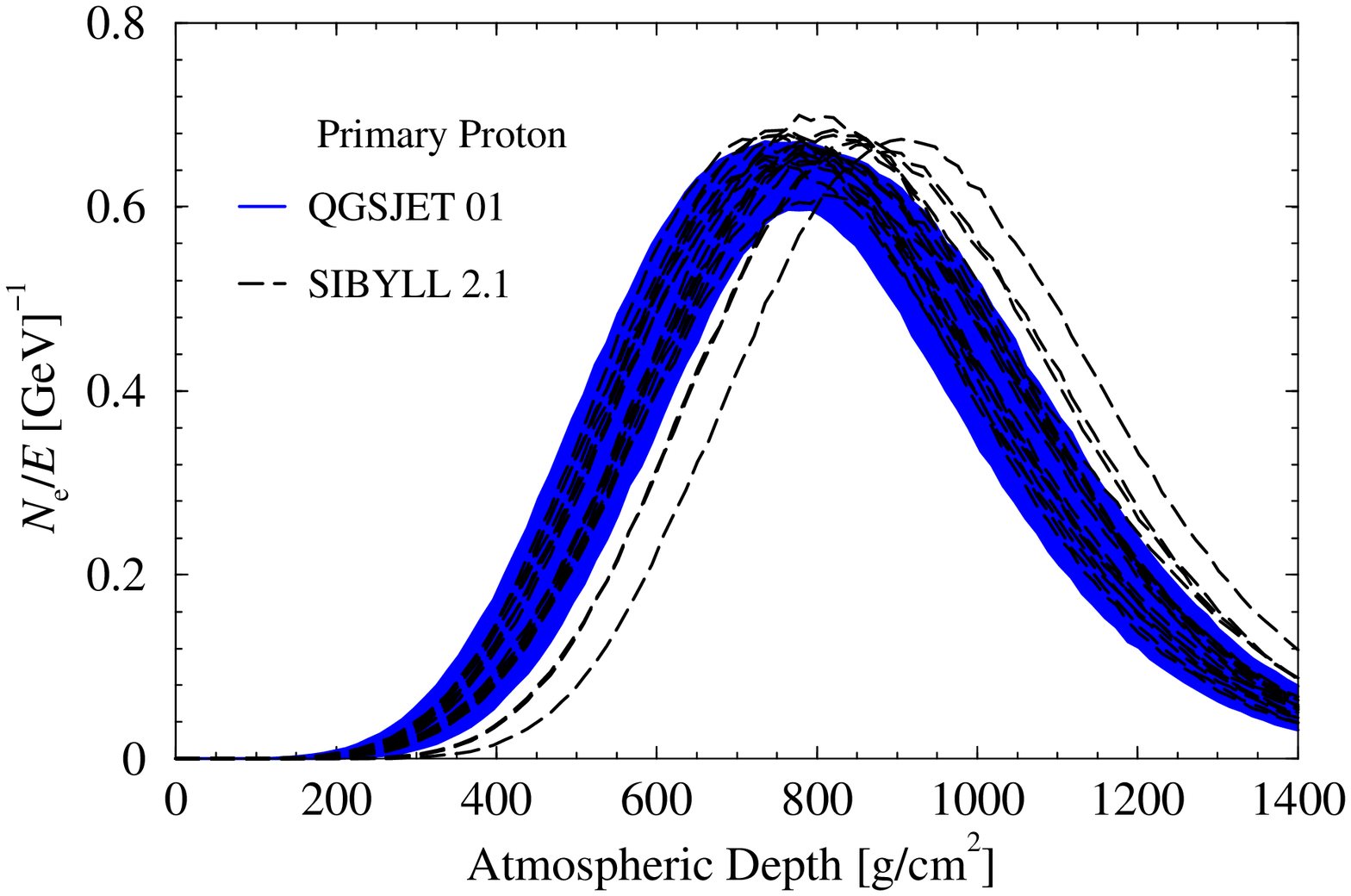}
\includegraphics[height=5.0cm]{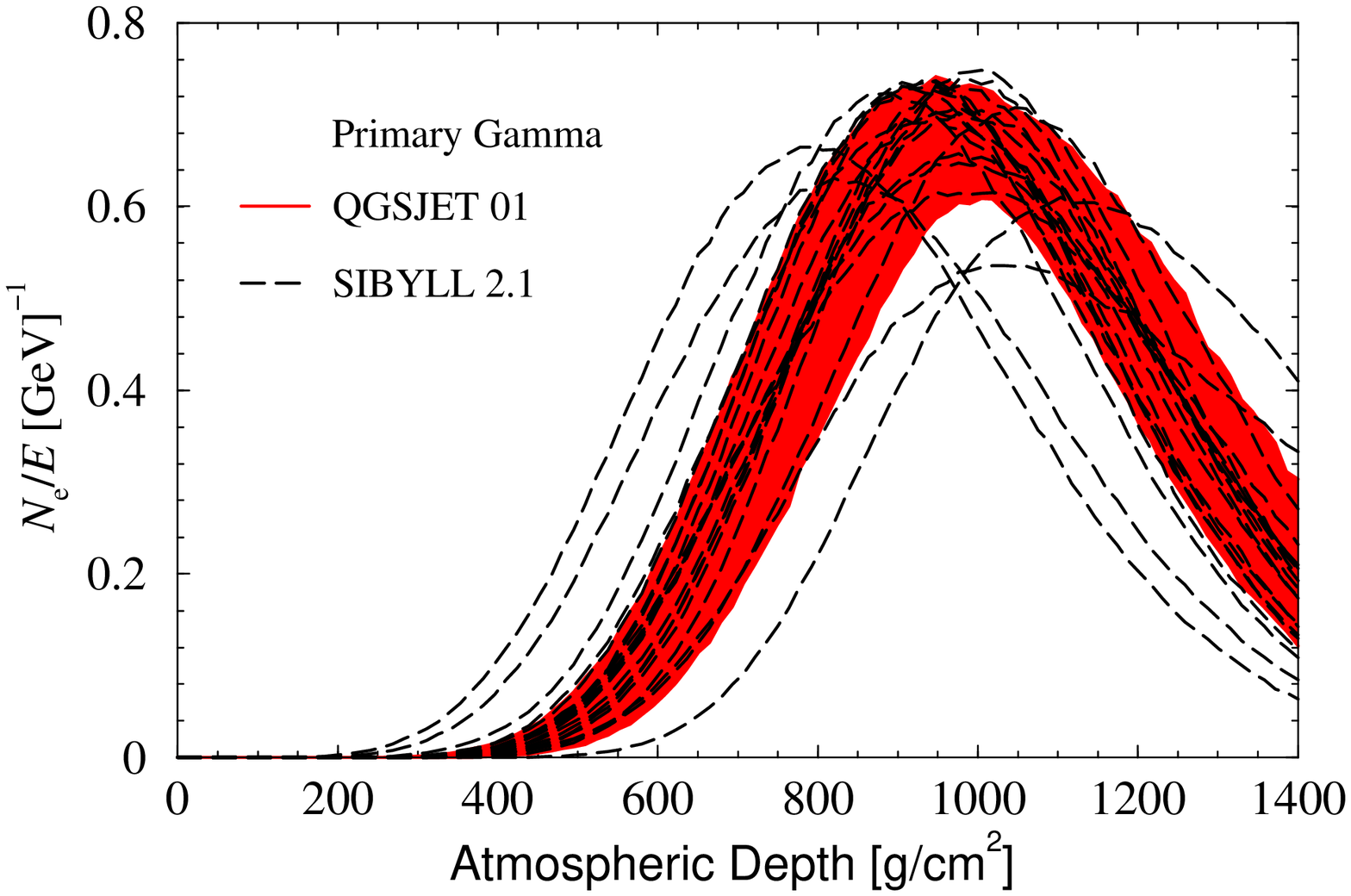}
\caption{\label{long_proton} Longitudinal profiles of $e^\pm$
illustrates the upper limit of 68\% of confidence level for 1000
QGSJET01 shower profiles (dashed area) with 20 random SIBYLL2.1
showers (dashed lines) simulated with SENECA scheme, for proton
(left panel) and gamma (right panel) induced showers at primary
energy of 10$^{19}$~eV.}
\end{figure}
\vspace{-0.5cm}
%
%
\section{Results and Comparisons}

Although the simulation of showers at fixed energies is not a very realistic application
we intend in the present contribution to compare quantitatively SENECA predictions for
both hadronic interaction models, providing few comparisons with CORSIKA results. Such
study can be useful to many experiments which use the fluorescence technique.

In order to make a simple comparison Fig.~\ref{long_proton}
illustrates the upper limit of 68\% of confidence level for 1000
QGSJET01 shower profiles (dashed area) with 20 random SIBYLL2.1
showers (dashed lines) simulated with SENECA scheme, for proton
(left panel) and gamma (right panel) induced showers at primary
energy of 10$^{19}$~eV. It is possible to verify a reasonable
difference between the longitudinal profiles obtained for both
hadronic interaction models. Such predictions reveal distinct
descriptions of the $X_{\mathrm{max}}$ fluctuations. Moreover, the
evolution of hadron-induced showers depends on the elasticity of
the interaction defined as the fraction of energy carried by the
leading secondary particle. The hadronic model SIBYLL2.1 predicts
a larger elasticity than QGSJET01 and by this reason the showers
penetrate more in the atmosphere, as can be seen in the left panel
of Fig.~\ref{long_proton}.
\vspace{-0.2cm}
\begin{figure}[ht]
\includegraphics[height=5.0cm]{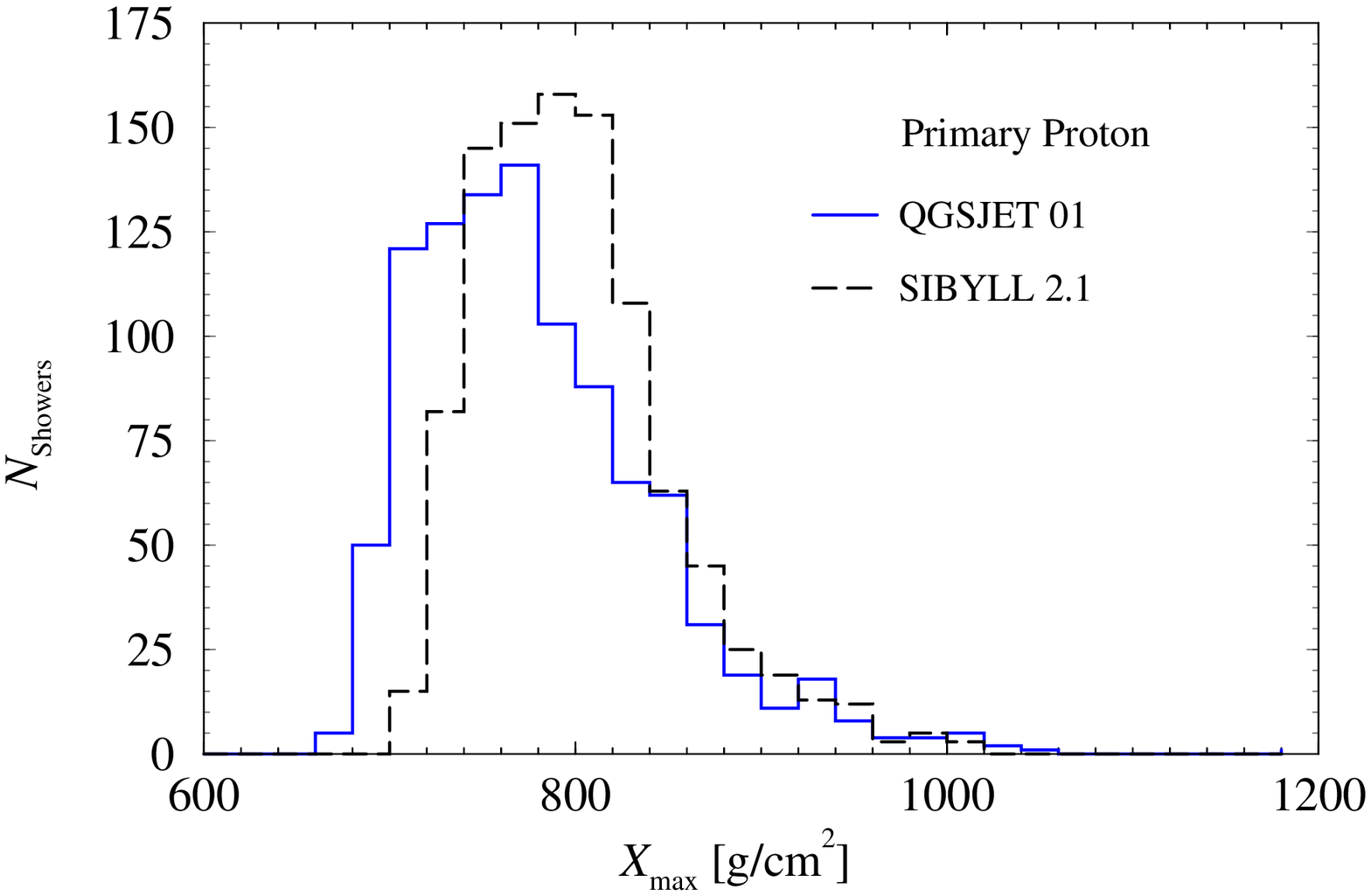}
\includegraphics[height=5.0cm]{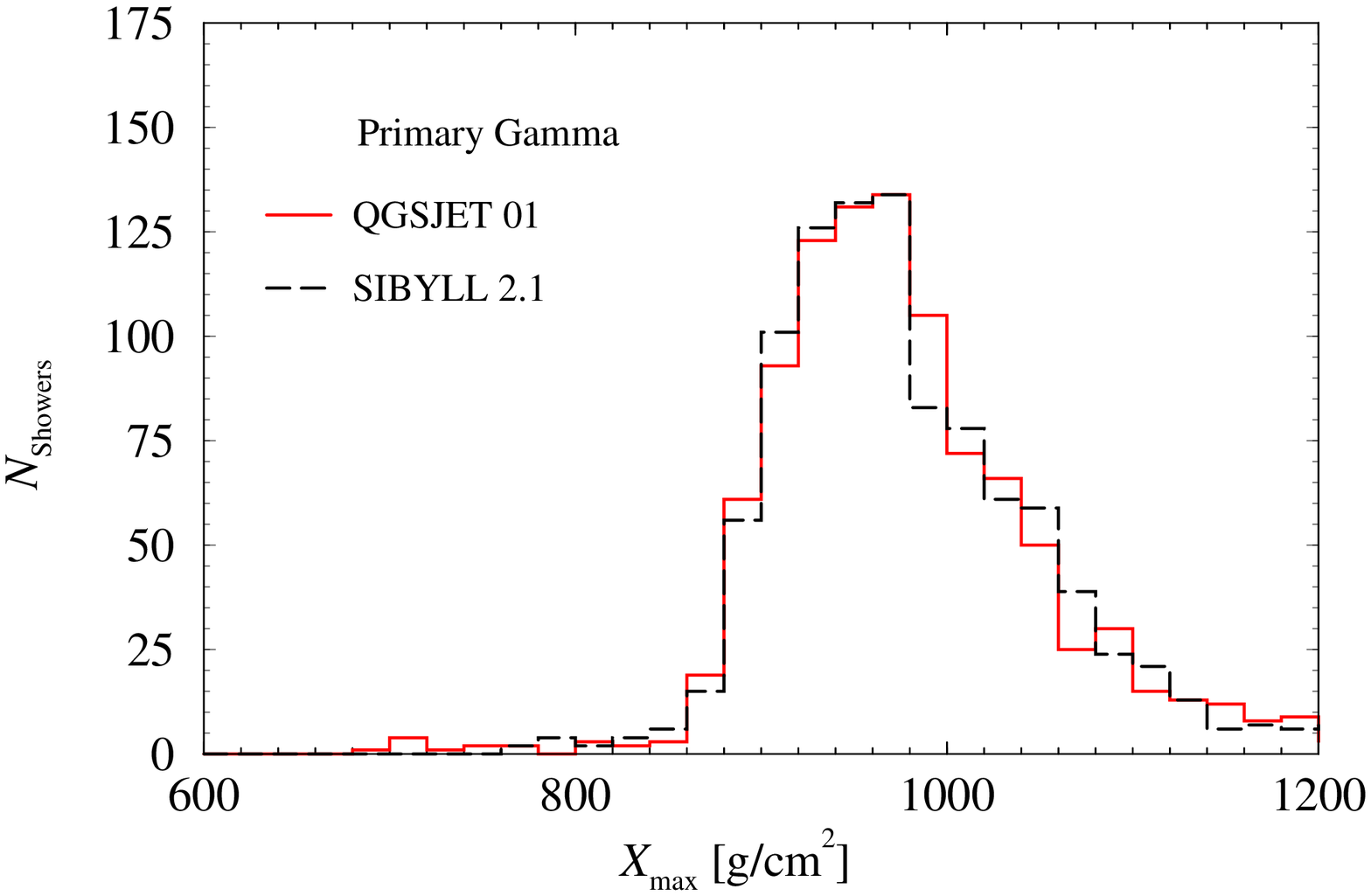}
\caption{\label{xmax_seneca} Distributions of the depth of maximum
air shower development shown for 1000 proton (left panel) and
gamma (right panel) showers at a particular primary energy of
10$^{19}$~eV, with incident zenith angle of 45$^{\circ}$,
generated by the hybrid technique with QGSJET01 and SIBYLL2.1
hadronic interaction models.}
\end{figure}
\vspace{-0.2cm}


Fig.~\ref{xmax_seneca} illustrates the potential of the
$X_{\mathrm {max}}$ distribution, generated with the hybrid
scheme, to distinguish possible primary signatures. In principle,
obtaining the values of $X_{\mathrm {max}}$ and/or $S_{\mathrm
{max}}$, by the fluorescence technique, and their respective
fluctuations, by Monte Carlo, one should be able to reconstruct
the shower energy and infer the identity of the primary cosmic
ray~\cite{Vitor}. The $X_{\mathrm {max}}$ distributions are
obtained for proton (left panel) and gamma (right panel) induced
showers, at a particular energy of 10$^{19}$~eV, with incident
zenith angle of 45$^{\circ}$, calculated with the hybrid technique
for both QGSJET01 and SIBYLL2.1 hadronic models.  Once more we can
clearly see the differences between the predicted average values
for hadronic shower observables calculated with QGSJET01 and
SIBYLL2.1 hadronic interaction models. Such differences can be
understood if we analyze the multiplicity. The charged particle
multiplicity is an important observable that measures how fast the
primary energy is dissipated into low energy sub-showers. The
QGSJET model predict a much higher multiplicity at high energies
when related with SIBYLL. For this reason, the $X_{\mathrm {max}}$
distribution obtained with the QGSJET model is shifted to lower
values and presents a wider distribution. For gamma showers, the
$X_{\mathrm {max}}$ distribution are almost identical.

As a final verification we show the correlation between
$S_{\mathrm {max}}$ and $X_{\mathrm {max}}$, which are important
longitudinal shower quantities on event reconstruction. We
compared SENECA results with CORSIKA predictions for gamma and
proton induced air showers, calculated with QGSJET01 hadronic
interaction model.
\begin{figure}[h]
\includegraphics[height=5.9cm]{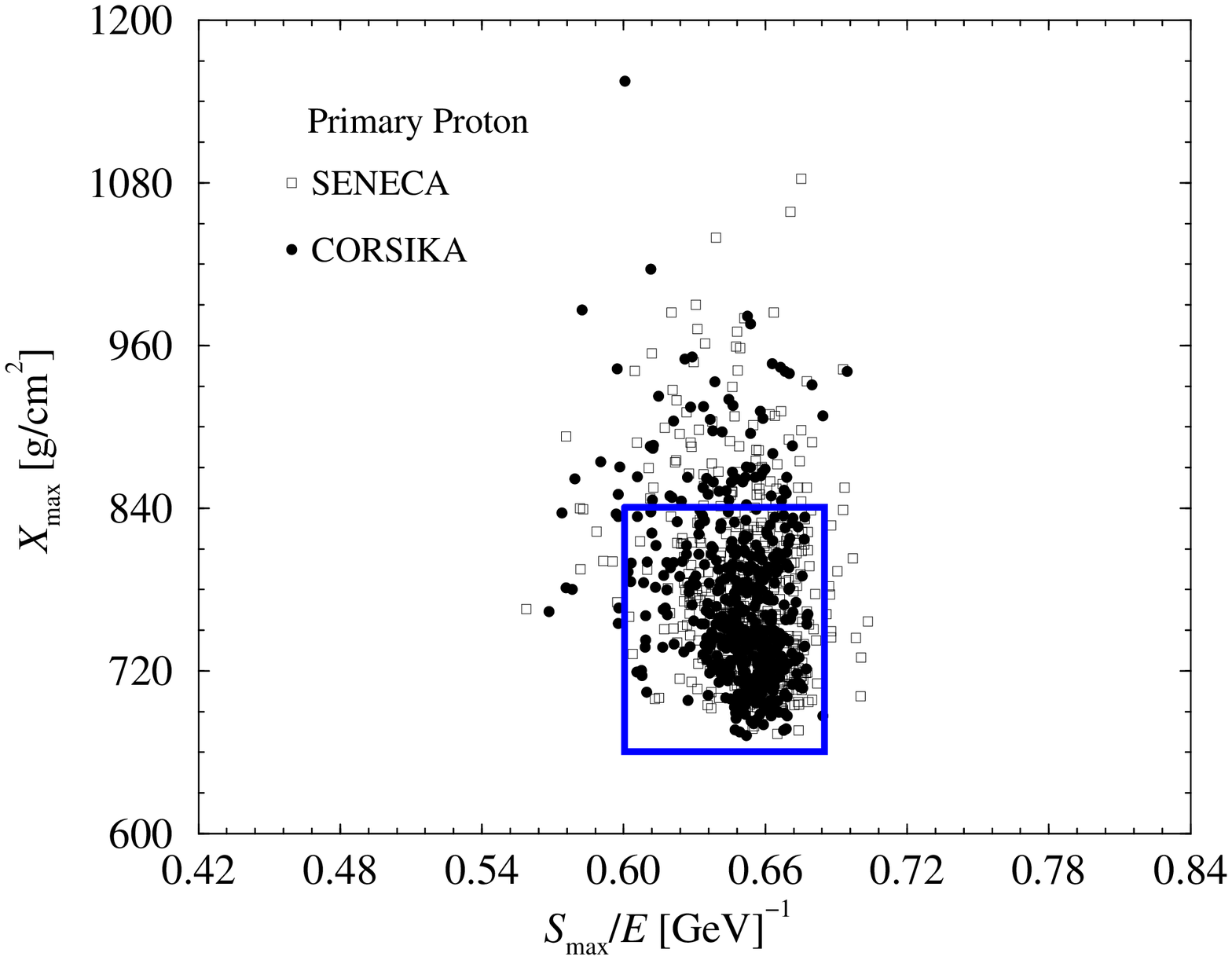}
\includegraphics[height=5.9cm]{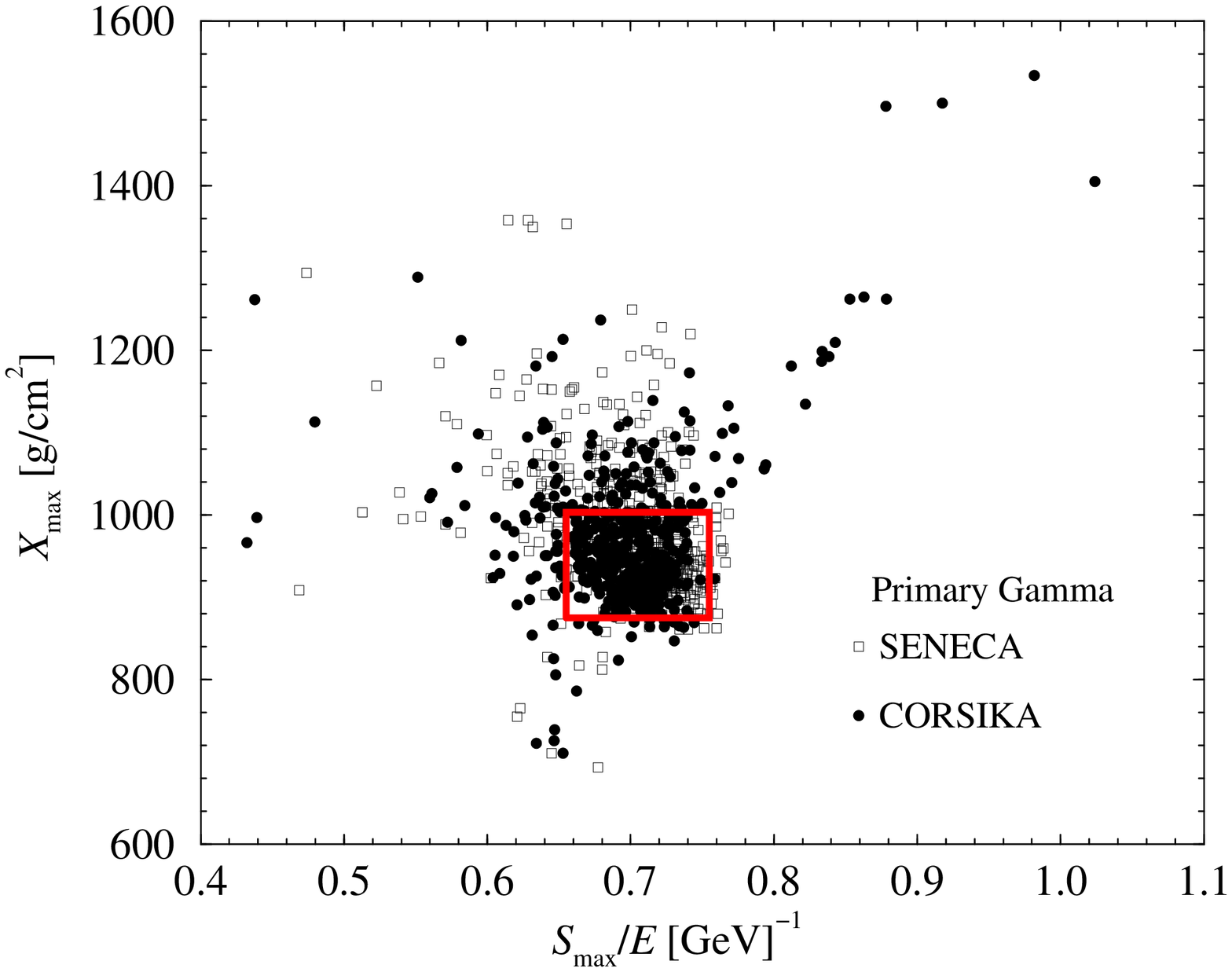}
\vspace{-0.2cm}
\caption{\label{smax_xmax} The correlation between $X_{\mathrm {max}}$ and $S_{\mathrm {max}}/E$
for proton (left) and gamma (right) induced showers at primary energy $10^{19}$~eV, at zenith angle
$\theta=45^\circ$, obtained with SENECA and CORSIKA codes. The full circles represent
500 showers simulated with CORSIKA, while the squares correspond to 500 showers generated with the hybrid method}
\end{figure}
\vspace{-0.1cm}
Fig.~\ref{smax_xmax} shows the correlation between these parameters for both simulation schemes.
We simulated 500 proton (left panel) and gamma (right panel) showers for each code, of energy
10$^{19}$~eV, incident zenith angle $\theta$=45$^{\circ}$ and free first interaction point. The
full circles (squares) denote the values for CORSIKA (SENECA). It is possible to verify the
existence of large fluctuations in gamma (right panel) ray showers at this particular primary
energy. The small box in the figure encloses, approximately, the highest density of correlation points for
both codes, showing very similar predictions: 61\% of the total number of events simulated by
SENECA, and 64\% of the total showers generated with CORSIKA. For the proton showers (left panel),
it is quite visible that the hybrid approach generates a more scattered distribution of showers,
with wider tales, than CORSIKA does. This is confirmed by counting the number of events inside
the small box shown in the figure. SENECA expectation amounts to 80\% of the proton showers
falling inside the box while CORSIKA expectation is 85\%.
The differences mentioned here are most, in principle, due to the air shower fluctuations
and are visibly model dependent.

\vspace{-0.2cm}
%
%
\section{Conclusions}

In the present work we analyzed the practical potential of SENECA, a very fast hybrid
tri-dimensional code, for the simulation of the longitudinal development of extensive
air showers at high energies.
The consistency of the SENECA scheme was tested and it proved to be very stable for
energies above $10^{18}$~eV. The results obtained by both codes agree well for the
analyzed quantities. Our careful analysis shows that many shower quantities discussed
here are strongly model depended. The undisputable bounty of SENECA is velocity which
(see~\cite{jao}), to say the least, is impressive over the primary shower energy simulated for this
contribution.

%
%
\section{Acknowledgments}

This paper was partially supported by the Brazilian Agencies CNPq and FAPESP.
Most of simulations presented here were carried on a Cluster Linux TDI, supported
by Laborat\'orio de Computa\c c\~ao Cient\'{\i}fica Avan\c cada at Universidade de
S\~ao Paulo.

\vspace{-0.5cm}
%
%

\end{document}